%% file: Manuscript_MPesceRollins.tex
\documentclass[preprint,3p,twocolumn,longtitle]{elsarticle}
\usepackage{amsmath}
\usepackage{graphicx}

\journal{Astroparticle Physics}

\begin{document}

\title{In-flight measurement of the absolute energy scale of the Fermi Large Area Telescope}

\include{AuthorList}

\begin{abstract}
The Large Area Telescope (LAT) on-board the Fermi Gamma-ray Space Telescope 
is a pair-conversion telescope designed to survey the gamma-ray sky from 
20~MeV to several hundreds of GeV. In this energy band there are no 
astronomical sources with sufficiently well known and sharp spectral features 
to allow an absolute calibration of the LAT energy scale.  However, the 
geomagnetic cutoff in the cosmic ray electron-plus-positron (CRE) spectrum in 
low earth orbit does provide such a spectral feature. The energy and spectral 
shape of this cutoff can be calculated with the aid of a numerical code tracing 
charged particles in the Earth's magnetic field. By comparing the cutoff value 
with that measured by the LAT in different geomagnetic positions, we have 
obtained several calibration points between $\sim$6 and $\sim$13~GeV with 
an estimated uncertainty of $\sim$2\%.  An energy calibration with such high 
accuracy reduces the systematic uncertainty in LAT measurements of, for 
example, the spectral cutoff in the emission from gamma ray pulsars.
\end{abstract}

\begin{keyword}
Cosmic-Rays \sep geomagnetic cutoff \sep absolute energy scale \sep Fermi Large
Area Telescope
\end{keyword}

\maketitle

\footnotetext[1]{Corresponding author}

\section{Introduction}\label{sec:intro}
The Large Area Telescope (LAT) is the primary instrument on the Fermi Gamma-ray
Space Telescope, which was launched on June 11, 2008. The LAT is a 4$\times$4 
modular array of identical towers, each comprised of a tracker/converter (TKR) 
and a calorimeter (CAL), surrounded by an anti-coincidence detector (ACD). Each
TKR module contains 18 layers of tungsten foil and paired silicon strip detector
$x$-$y$ tracking planes. Each CAL module contains a hodoscopic array of 
96 CsI(Tl) scintillating crystals 8.6 radiation lengths deep. High energy gamma
rays predominantly convert in the TKR into electron-positron pairs, and the 
energy of the developing electromagnetic shower is measured in the CAL. 
To cover the energy range of the instrument the individual CAL crystals must 
be able to accurately measure energy deposits between 2~MeV and 60~GeV.  To 
accomplish this, each CAL crystal can measure the deposited energy in four 
different overlapping energy ranges, namely 2--100~MeV, 2~MeV--1~GeV, 
30~MeV--7~GeV, and 30~MeV--70~GeV.

The process of converting the scintillation light measured in each individual 
CAL crystal to an energy of the incident photon can be roughly divided into 
two phases. The first phase is the calculation of the
energy deposition in each crystal (hereafter, the ``crystal energy"), which 
relies on calibration constants derived from the measured signals from 
sea-level muons and on-orbit cosmic ray protons together with constants derived
from an on-orbit electronic charge-injection system. The second phase is event
reconstruction, which is comprised of a set of algorithms that account for 
energy deposited in the TKR, leakage of the shower out the back and sides of 
the CAL, and energy lost in gaps between the towers. We developed the energy 
reconstruction algorithms from 
a detailed instrument model and Monte Carlo simulation based on 
GEANT4~\cite{latpaper}. 

To verify the fidelity of the instrument model and Monte Carlo simulation, 
we performed an extensive beam test campaign in
2006 at CERN and the GSI heavy ion accelerator laboratory. To eliminate the 
handling and shipping risks to the flight detector subsystems, and to reduce 
cost and schedule impacts, these tests were not done on the full LAT but on a 
Calibration 
Unit (CU), which consisted of two fully-populated flight-spare TKR modules 
and three fully-populated flight-spare CAL modules. A more detailed description
of the beam test as well as the energy reconstruction algorithms will be 
presented in section~\ref{sec:beamtest}.

To perform an in-flight verification of the absolute 
energy scale of the LAT it is necessary to find an astrophysical
source with a spectral feature whose absolute energy peak and shape are
well known. Because sources in the LAT 
energy band exhibit smooth spectral forms, it is not trivial to satisfy this 
requirement. Furthermore, the LAT is far more sensitive than any previous 
gamma-ray telescope, so it is not possible to use existing measurements of 
celestial sources to perform this calibration. The LAT has shown that 
gamma-ray pulsars typically exhibit spectral cutoffs in the 1--10~GeV 
range~\cite{PulsarCat} and, for bright pulsars, the spectral cutoff energy 
can be measured with statistical uncertainty approaching 1\% \cite{Vela1yr}.  
If the systematic uncertainty on the energy scale in this 
energy range can be reduced to the few-percent level, precise measurements of 
spectral cutoff can address detailed questions of the emission physics and 
geometry in pulsars~\cite{Vela1yr}. 
In the orbit of the Fermi observatory 
(inclination of 25.6$^{\circ}$ and altitude of 565 km), the geomagnetic cutoff 
in the cosmic-ray electron plus positron (CRE) spectrum can serve as such 
a source in the range between $\sim$6 and $\sim$13~GeV. Electrons and positrons
are a good calibration source because they interact in the LAT CAL in the
same way that photons do: they both produce electromagnetic showers.
The energy measurement bias coming from the slight differences between CRE 
and photons interactions in the LAT is much less than 1\%~\cite{fullpaper}.
The capabilities of the LAT to measure the CRE spectrum have been well 
demonstrated~\cite{fullpaper,prl}. The cutoff rigidity\footnote{Rigidity is defined as the particle momentum divided by its charge. The cutoff we are measuring in the CRE spectrum is not the vertical cutoff but rather averaged over all angles.} 
can be predicted by numerically tracing particle trajectories in the Earth's 
magnetic field, and the comparison of the predicted and measured values
provides the opportunity to perform this validation.

In order to measure the deviation between the reconstructed electron spectrum 
and the calculated value as a 
function of energy, this analysis is performed in several McIlwain L intervals.
The McIlwain L-parameter~\cite{walt} is a parameter 
describing a set of the Earth's magnetic field lines, in particular those
which cross the Earth's magnetic equator at a 
number of Earth-radii equal to the L-value.  Magnetically equivalent
positions (from the standpoint of the incoming charged particle) around the 
world will by definition have the same McIlwain L values,
therefore making this parameter particularly convenient for characterizing 
cutoff rigidities~\cite{smartshea}. The orbital inclination of the Fermi orbit 
fixes the range of McIlwain L values accessible for this analysis to 
1.00--1.72.

\section{LAT energy calibration}\label{sec:beamtest}

The absolute energy scale of the LAT is defined by comparing the signals in 
the CAL crystals with the amount of energy a Monte Carlo simulation indicates 
should be deposited by on-orbit relativistic protons. The details of the energy
calibration, referred to as the "proton inter-range calibration", are given 
in~\cite{onorbitcalib}. To perform this calibration, we first correct the 
observed CAL 
signals for electronic non-linearities (as measured by an electronic 
charge-injection process) and for position-dependent scintillation response 
(as measured by a direct calibration with sea-level muons and on-orbit with 
protons). We then compare the corrected CAL signals with the distribution of 
deposited energies predicted by a GEANT4 simulation of the on-orbit spectrum 
of primary cosmic-ray protons passing through the LAT. This calibrates the 
highest-gain, lowest-energy range of the CAL readout. The remaining gain 
ranges~\cite{latpaper} of the CAL readout are calibrated by enforcing that 
adjacent gain ranges give the same measured energy in the regions of energy 
space in which they overlap~\cite{onorbitcalib}. 

The Galactic cosmic-ray (GCR) element abundance peaks could in principle be 
used to calibrate the higher energy ranges of the LAT; however, the 
scintillation efficiency of heavy ions in CsI(Tl) is not identical to that of 
electromagnetic showers by an element-dependent factor that is not well known 
or measured.
We measured the relative scintillation efficiencies of 
sub-relativistic ions in CsI(Tl) crystals in a series of beam 
tests~\cite{LOTTGsi},
but we do not have sufficient confidence that these efficiencies apply at 
relativistic energies.  Thus, we are unable to relate the observed signals 
from GCR element peaks to electromagnetic shower energy depositions accurately 
enough to calibrate the LAT energy scale. 

The geomagnetic cutoff in the CRE spectrum (measured in the energy range 
$\sim$6 to $\sim$13 GeV) is a good alternative source to calibrate the higher 
energy ranges of the CAL readout. At these energies, the maximum energy per 
crystal is of the order of $\sim$1 GeV, and will be read out in the 
higher energy ranges of the CAL. As a reference, 1 GeV of energy per crystal is 
equivalent to 100 times what a minimum ionizioning particle releases per 
crystal.

The LAT CU was also calibrated using sea-level cosmic-ray muons, but the 
absolute energy scale was then cross-checked against calibrated beam lines at 
CERN, showing an energy deposit systematically larger than expected.
Such a direct calibration relies on the knowledge of the incoming beam energy 
(determined with a $1\%$ accuracy), the geometry of the beam line and the
CU detector, the Monte Carlo of the electromagnetic shower development within 
the detector, as well as on a good control of the environmental effects 
that might influence the CU response (temperature, humidity, exceedingly high 
particle rates).

The energy reconstruction of the LAT (as well as of the CU) is based on
three different algorithms: a parametric correction based on the barycenter of 
the shower, a fit to the shower profile taking into account the longitudinal
and transverse development of the shower, and a maximum likelihood fit based
on the correlation between the total deposited energy, the energy deposited in
the last layer of the CAL and the number of tracker hits. For each event,
the best energy reconstruction method is selected by means of a classification
tree analysis described in~\cite{latpaper}. We would like to emphasize that
at the energies considered in this analysis (i.e. between $\sim$6 and $\sim$
13 GeV) the correction factors for losses due to leakage out of the detector
are reasonably small (on average of the order of 30\%) when compared to higher 
energies (of the order of 50\% for 100's of GeV) and therefore the method
described here is testing a mixture of the crystal energy calibration and
the leakage corrections.

Following the beam test campaign, the GEANT4 simulation of the CU was updated 
to reflect the best available description of the CU detector, the beam line, 
and the particle interactions. Most notably, the routines describing the 
Landau-Pomeranchuk-Migdal effect (LPM~\cite{LMPEffect}) in electromagnetic 
shower developments were updated after finding that initial GEANT4 
implementations were not providing a satisfactory description of the shower 
longitudinal development. The energy resolution for all three 
available energy reconstruction methods was measured to be consistent with 
expectations. However despite these improvements, the reconstructed energy in 
the CU was consistently higher than the beam energy by $\sim$9\%, on 
average, with further fluctuations of $\sim$5\%, depending on incoming beam 
energy (5--282~GeV), angle (0--60~degrees) and position of the crystal 
within the shower~\cite{beamtest,fullpaper}.

The energy scale factor derived from the direct CU calibration 
was not applied to the LAT because of the differences in the beam test and 
on-orbit environment, particularly the distribution and rate of the particles 
incident on the calorimeter, and the temperature variations for the CU during 
the data taking, which were considered responsible for the different LAT and CU 
energy scales.

\section{Particle Tracing}\label{sec:tracer}
Geomagnetic cutoff rigidities can be obtained by tracing 
cosmic-ray trajectories in a model of the Earth's magnetic 
field. The standard mathematical description of the Earth's magnetic field
is given by the International Geomagnetic Reference Field models
(IGRF)~\cite{IGRFPaper}. These models consist of the Gauss coefficients 
defining 
the spherical harmonic expansion of the geomagnetic potential 
up to a given order. In the IGRF-11 model, used for this analysis, the maximum 
multipole moment of the expansion is the 13$^{\rm th}$. The IGRF model is 
computed by the participating members of the International Association of 
Geomagnetism and Aeronomy (IAGA) Working Group V-MOD~\cite{IGRFPaper} which is 
supported by the organizations involved in operating magnetic survey 
satellites, observatories, magnetic survey programs and World Data Centers.
IGRF-11 provides a definitive main field model for epoch 2005, a main field 
model for 2010, and a linear predictive secular variation model for 
2010--2015. We use the particle trajectory
tracing code (hereafter tracer) developed by Smart and Shea~\cite{smartshea} 
and the IGRF-11 in this analysis to measure the geomagnetic cutoff.

\begin{figure}[h!]
\begin{center}
\includegraphics[width=\linewidth]{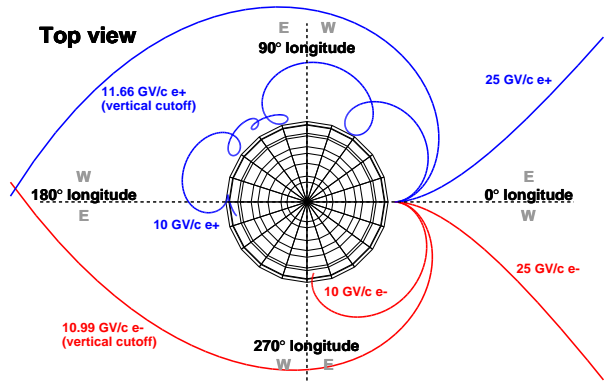}
\end{center}
\caption{Illustration of $e^-$ and $e^+$ of varying 
  energies traced in the Earth's magnetic field from a point at longitude 
  and latitude (0,0) and 565 km above the Earth's surface and viewed from the
  North Pole. The vertical cutoff
  is taken to be the lowest rigidity which the particle needs to escape when 
  traced vertically through the Earth's magnetic field. As the rigidity of 
  the particle decreases the amount of geomagnetic bending increases and 
  when the trajectory intersects the Earth it is taken to be forbidden 
  (i.e of secondary origin). The cutoff values are different for positively 
  and negatively charged particles because the Earth's magnetic
  field is not a perfect dipole. 
  Particles labeled as escaped are taken to be of Galactic origin.
  Earth and trajectories are to scale.}
\label{fig:trajectories}
\end{figure}

Since it is difficult to trace the trajectory of an incoming particle
through the magnetic field and expect it to intersect the precise location
desired, it is more efficient to calculate the trajectory in 
the reverse direction. Therefore the starting point of the trajectory is 
given by the geographic coordinates, altitude and orientation of the spacecraft
and the trajectories are propagated using IGRF-11~\cite{IGRFPaper}. 

At each of a grid of locations covered by the Fermi orbit, evaluated every 15
seconds from Aug 2008 through July 2009, we simulated an
ensemble of test particles. We selected energies of the test particles according
to the power-law spectrum measured by the LAT~\cite{fullpaper}, and generated 
both electrons and positrons in the abundance ratio measured by PAMELA~\cite{pamela}. We use the tracer code to determine for each test particle whether it 
could have originated from outside the geomagnetosphere. Trajectories that 
eventually intersect the Earth's atmosphere\footnote{Distance taken to be 20 km from the Earth's surface.} are rejected and labeled as forbidden (secondaries 
hereafter). Trajectories 
that reach 20 Earth radii are accepted as escaped (or of Galactic origin, 
primaries hereafter). 
Figure~\ref{fig:trajectories} illustrates a few sample trajectories, both 
forbidden and allowed.

\begin{figure}[ht!]
\begin{center}
\includegraphics[angle=90,width=\linewidth]{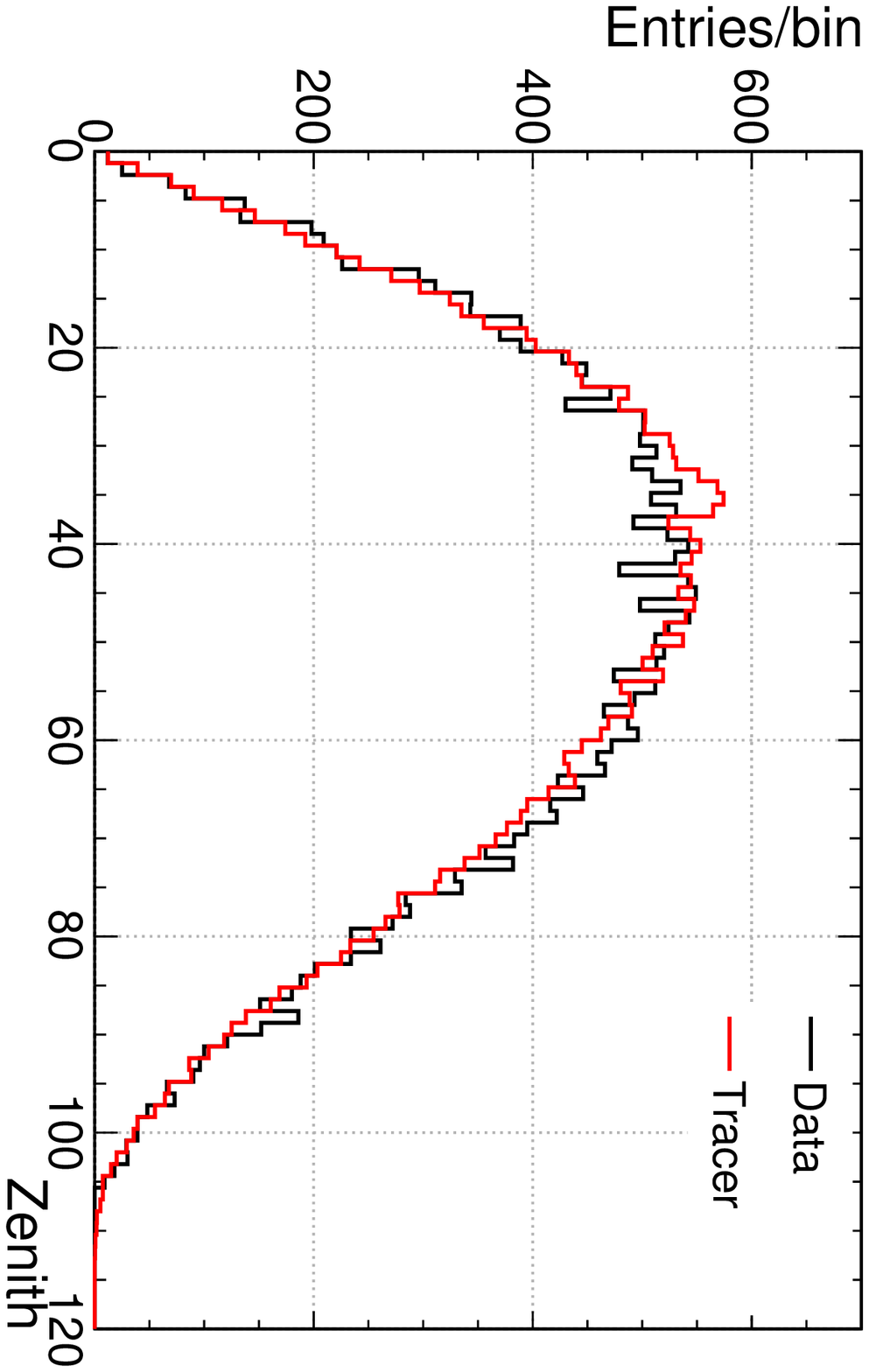}
\includegraphics[angle=90,width=\linewidth]{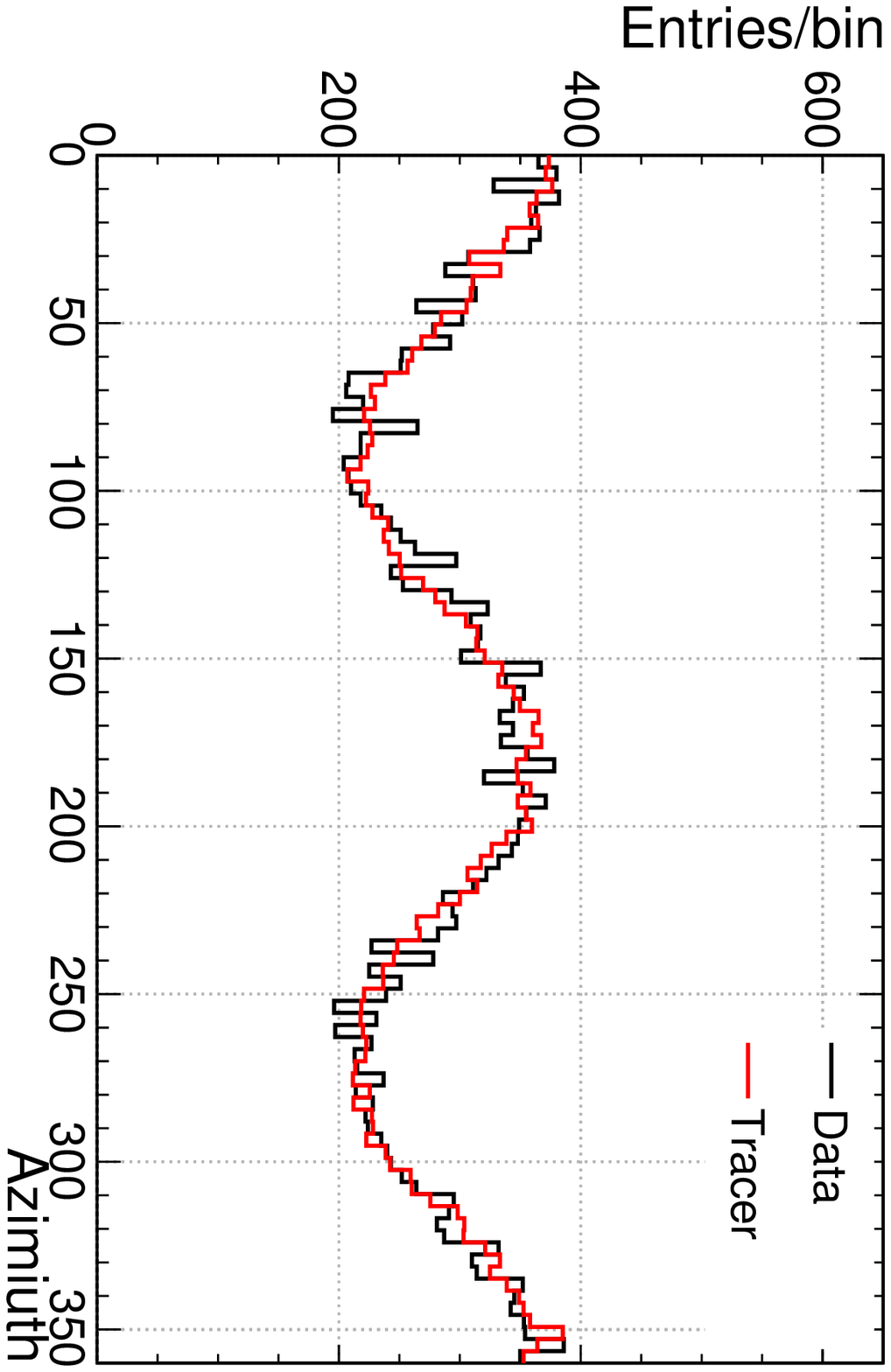}
 \end{center}
\caption{The reconstructed angle with respect to local zenith (upper panel)
and azimuth (lower panel) for data and tracer. In the lower 
panel, north is at 0$^{\circ}$, east at 90$^{\circ}$, 
south at 180$^{\circ}$ and west at 270$^{\circ}$. The LAT acceptance 
for electrons and positrons as well as the rocking profile have been convolved 
in these distributions and influence their shape. Both are averaged
over the orbit and for energies greater than 30~GeV in order to compare the 
distribution of primary CREs. There is an overall good agreement between 
data and tracer.}
\label{fig:angularDist}
\end{figure}

We selected electron and positron data from the first year of Fermi LAT data 
taking (Aug 2008 through July 2009) using the methods described in 
\cite{fullpaper} for the 100~MeV to 100~GeV energy range. In 
figure~\ref{fig:angularDist} are the distributions of the reconstructed 
angle with respect to local zenith (upper panel) and azimuth (lower panel) 
for the flight data (black) and
for the tracer output (red). The overall agreement between flight data and
the tracer output is very good, demonstrating that the angular distributions 
have been well described.

\section{Analysis}\label{sec:analysis}
\begin{figure*}[ht!]
\begin{center}
\includegraphics[angle=90,width=0.49\linewidth]{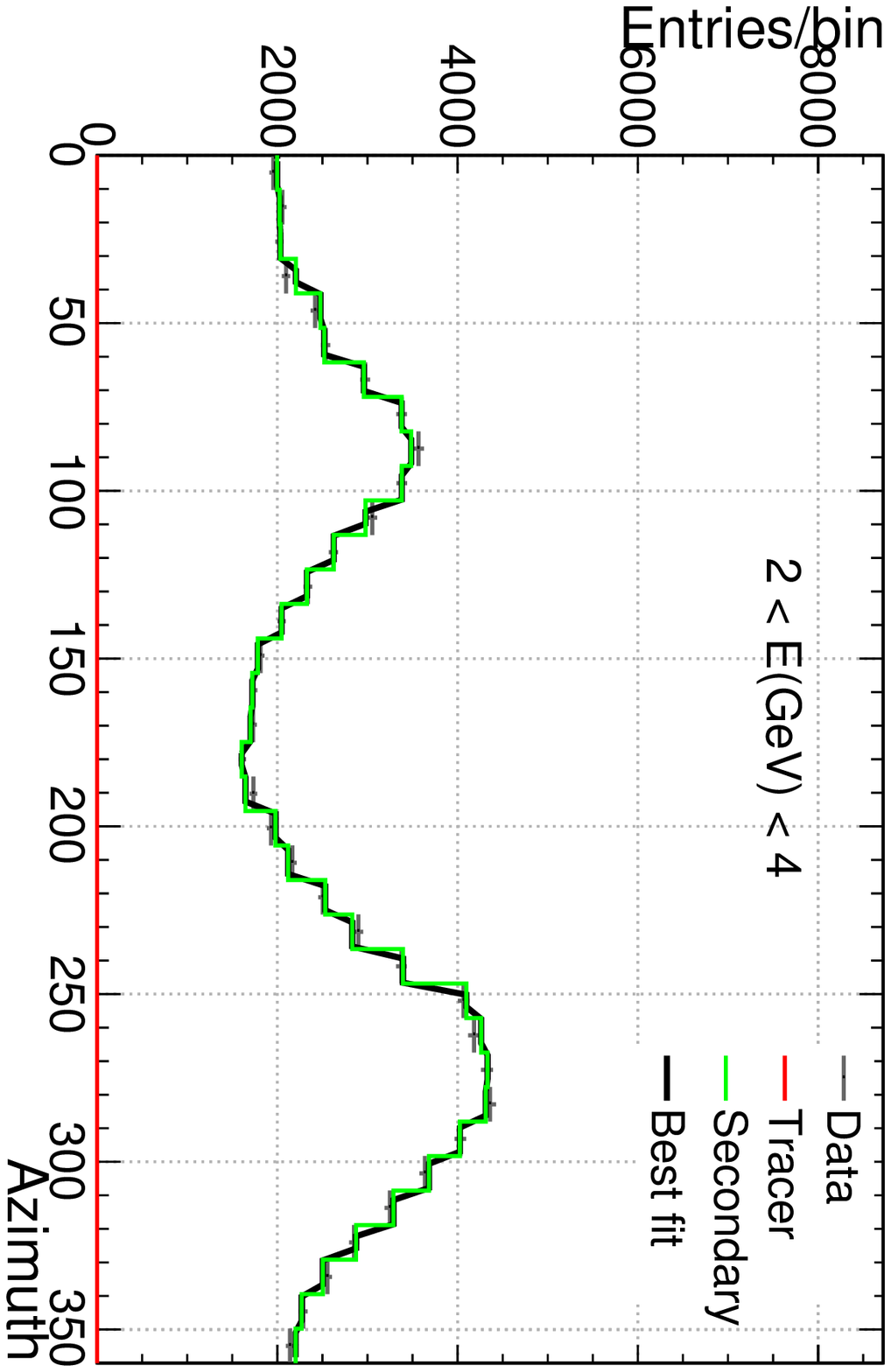}
\includegraphics[angle=90,width=0.49\linewidth]{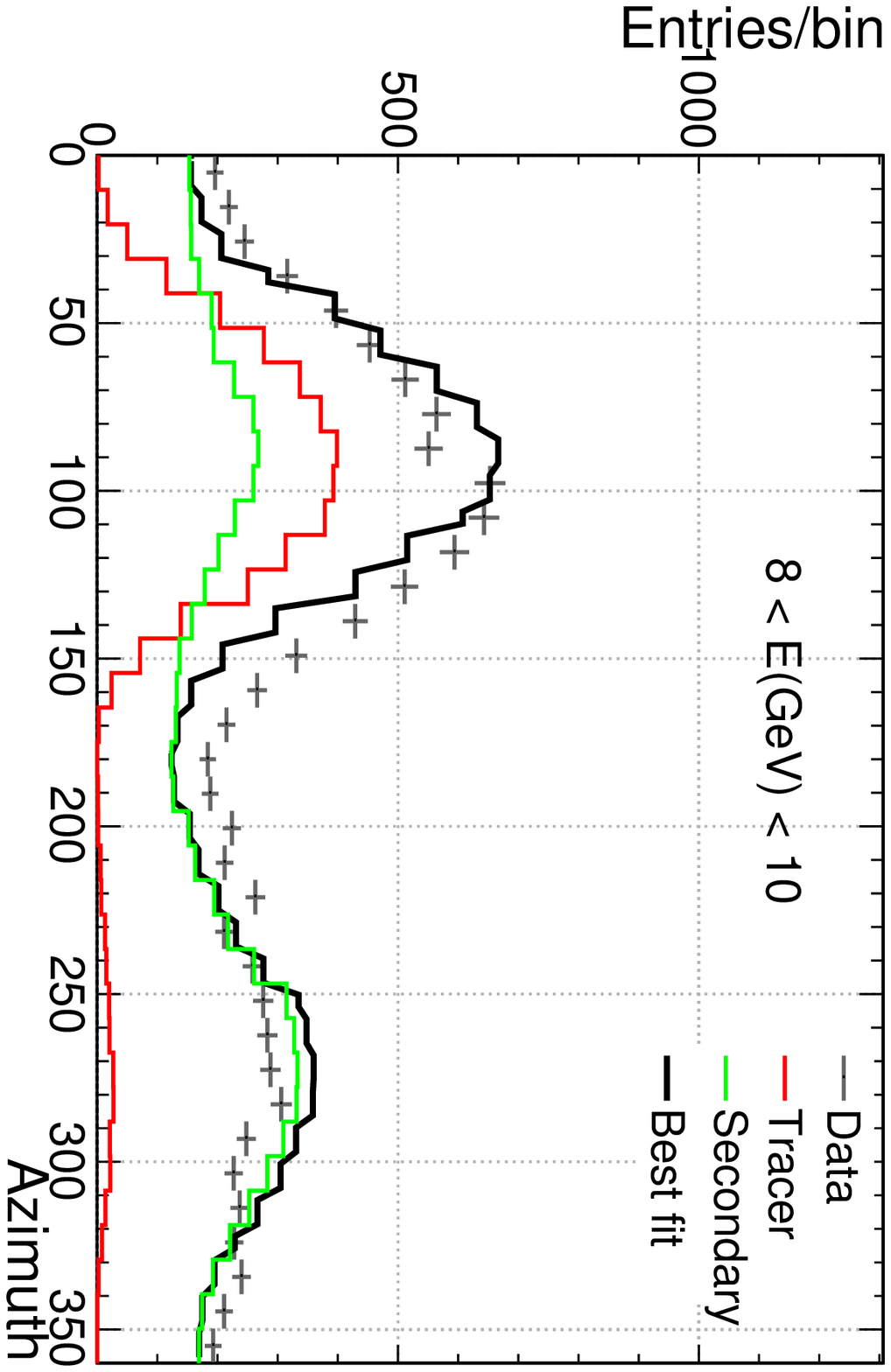}
\includegraphics[angle=90,width=0.49\linewidth]{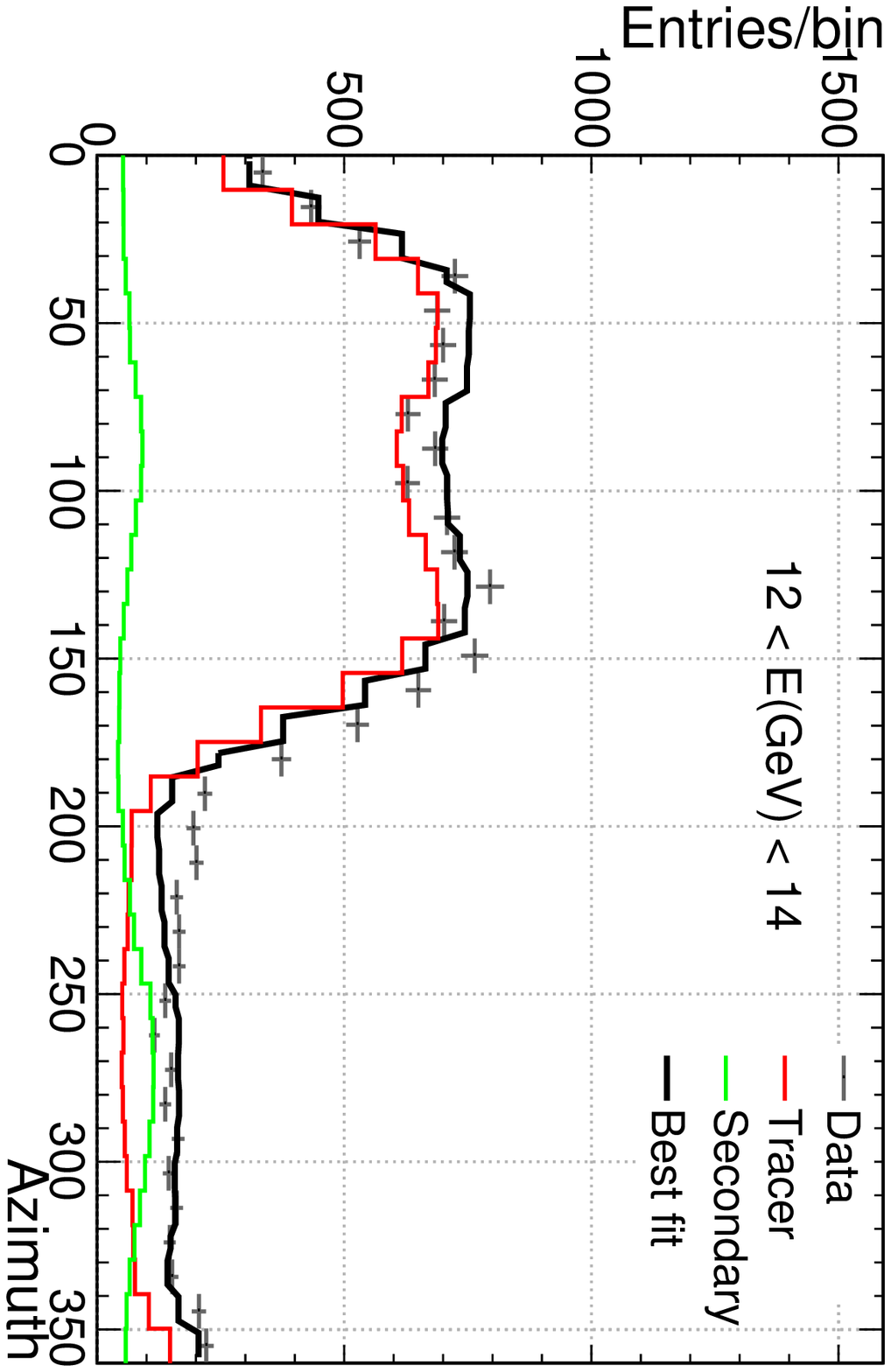}
\includegraphics[angle=90,width=0.49\linewidth]{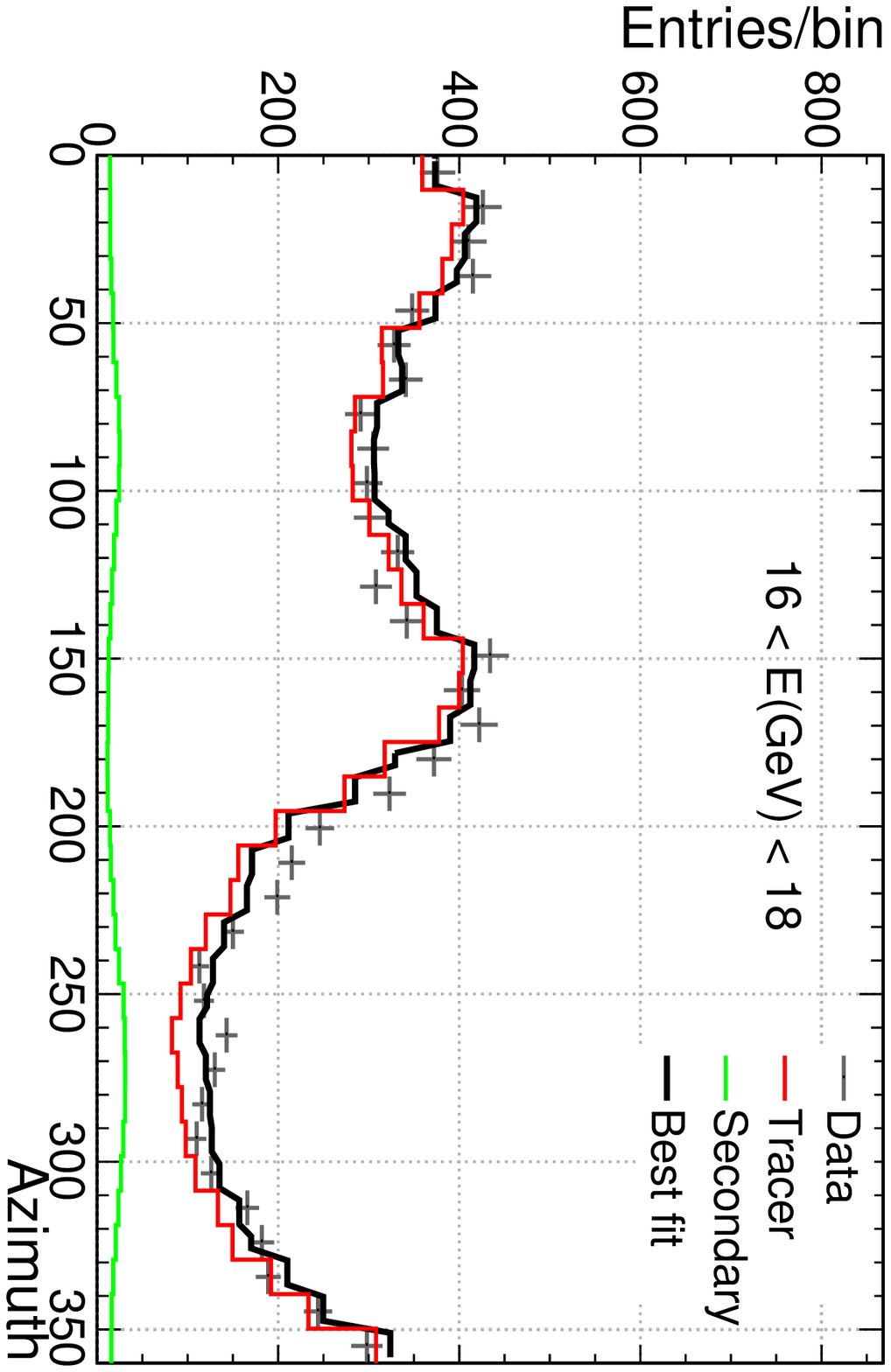}
\end{center}
\caption{Azimuthal distribution for flight data, tracer 
and secondaries in the McIlwain L interval 1.0~$<$~L~$<$~1.14. 
The black distribution is the linear combination of tracer and secondaries 
which 
best fits the data. Each panel depicts the distributions in a given energy 
interval (labeled in the panel). In the upper left panel, where the secondary
population is dominant, a maximum is evident in the westward direction where
the secondary positrons are dominant. With
increasing energy, the primary population gradually increases in the eastward
direction. The LAT acceptance and rocking profile have been convolved, which 
in turn effects the overall shape of the azimuth distribution. The cutoff 
rigidity in this McIlwain L range is $\sim$13~GeV. The fraction of primaries 
evaluated from this template fitting is shown in figure~\ref{fig:albedoVsprime}}
\label{fig:primeFrac}
\end{figure*}
\subsection{Estimating the fraction of primary Cosmic Ray electrons}
\label{sec:primefrac}
The population of CREs in low Earth orbit is a mixture of primary and 
secondary cosmic rays, where primary implies of extraterrestrial origin and 
secondary 
implies a mixture of splash and re-entrant particles produced in the interaction
of primaries in the Earth's atmosphere.
The trajectories of the secondary particles are extremely difficult to 
simulate reliably. By definition our tracing 
code only provides information on the escaped (i.e. primary) particles.
As a consequence it is necessary to estimate the fraction of secondary 
particles, as a function of energy, from the flight data.
This constitutes one of the most delicate aspects of the analysis 
because the cutoff rigidity is found by fitting the spectrum (as will be
described in detail in section~\ref{sec:cutoffrigidity}) and its shape
is influenced by the fraction of secondaries. 

\begin{figure}[h!]
\begin{center}
\includegraphics[angle=90,width=\linewidth]{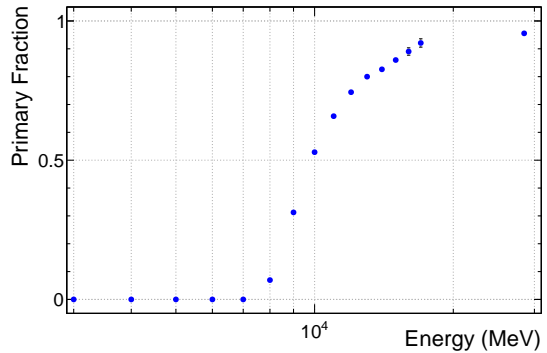}
\end{center}
\caption{Fraction of primary electrons and positrons 
as a function of energy for the McIlwain L interval 1.00~$<$~L~$<$~1.14}
\label{fig:albedoVsprime}
\end{figure}

The azimuthal distribution in Earth centered coordinates of the secondary 
population is different from that of the 
primary particles. We can exploit this fact and perform 
a template fit to identify the fraction of each population. 
To perform this task it is first necessary to choose the appropriate templates 
to describe the populations. It is safe to assume that 
the population at low energy in flight data (E$<<$E$_{\rm c}$, where E$_{\rm c}$
is the energy corresponding to the cutoff rigidity) is predominately
composed of secondary particles, and can therefore be used as a template.
The output from the tracer code provides the template for the primaries.
Figure~\ref{fig:primeFrac} shows an example of the
template fitting performed to estimate the fraction of primaries 
in the interval 1.0 $<$ McIlwain L $<$ 1.14. The gradual transition from a
pure sample of secondary particles (top left panel, for energies between 2 and 
4~GeV) to that of only primaries (bottom right panel, for energies between
16 and 18~GeV) is evident. For reference, the cutoff energy in this McIlwain 
L interval
is $\sim$13~GeV. For simplicity, only four energy intervals are shown.
However the analysis has been performed in 16 overlapping energy intervals.
The resulting value for the fraction of primaries as a
function of energy for this same McIlwain L interval is shown in figure~\ref{fig:albedoVsprime}.

\subsection{Measuring the cutoff rigidity}\label{sec:cutoffrigidity}
 The primary $e^- + e^+$ spectrum can be parameterized by~\cite{fullpaper}:
\begin{equation}
 \frac{dN(E)}{dE} =  \frac{cE^{-\Gamma}}{(1+(E/E_c)^{-6})}
\label{eq:fitfunction}
\end{equation}
Where $\Gamma$ is the spectral index and $E_c$ is the 
energy corresponding to the cutoff rigidity. We fit the primary CRE spectrum 
(both data
and tracer) with equation~\ref{eq:fitfunction} to get the value of the 
cutoff rigidity.

To reconstruct the CRE primary spectrum it is first necessary to remove
the secondary population from the count spectrum and then correct for the 
hadronic contamination. The residual contamination from hadrons, $h$, is 
estimated by applying the
CRE event selection to the on-orbit simulation used by the 
Fermi LAT collaboration. We found that this residual hadron contamination 
ranges between $\sim$15\% below the cutoff and $\sim$5\% above with an 
estimated uncertainty on the absolute hadron flux of 20\%. The on-orbit 
simulation is based on data from CR 
experiments and includes all the components
of the Galactic cosmic rays as well as the re-entrant and splash Earth
albedo particles. A detailed description of this simulation can be found 
in~\cite{fullpaper}. The residual contamination due to secondaries, $s$,
is estimated as described in section~\ref{sec:primefrac}. The count 
spectrum is multiplied by
the purity factor defined as $P.F = (1-h)\cdot(1-s)$ in each energy bin.
This background-subtracted count spectrum is divided by the effective
geometry factor (EGF)~\cite{fullpaper} and the width of the energy interval,
to obtain the final spectrum.

The tracer count spectra are reconstructed by requiring
the escaped condition and dividing by the width of the energy interval. To 
properly compare the cutoff rigidities, 
the tracer spectra have been binned in measured energy (i.e., they have been 
convolved using the energy 
resolution of the Fermi LAT for electrons and positrons~\cite{fullpaper}).
No efficiency correction is required for the tracer spectra.
To measure the spectra in McIlwain L intervals 
an extra cut specifying the interval is applied to both data and tracer. 

\begin{figure}[h!]
\begin{center}
\includegraphics[angle=90,width=\linewidth]{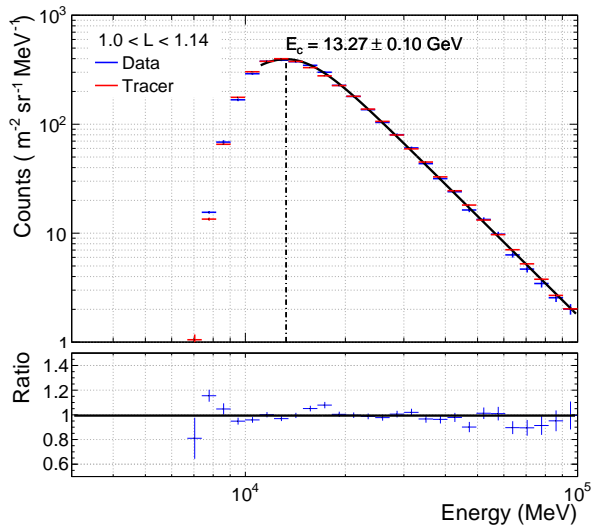}
\end{center}
\caption{Energy spectrum of data and tracer for the McIlwain L
interval 1.0$<$L$<$1.14. The black line is a fit to the data using equation 
\ref{eq:fitfunction} and the resulting cutoff energy has a value of
13.27$\pm$0.10~GeV, and is indicated by the dot dashed line in the
figure. The lower panel depicts the ratio of the two spectra.}
\label{fig:spectrumGraph}
\end{figure}
A comparison between the counts spectra measured from flight data and from
the tracer output is shown in figure \ref{fig:spectrumGraph}. The shapes of the
spectra are in good agreement, indicated by the ratio of the two 
depicted in the lower panel of the figure. 
Both the reconstruction and fitting procedures are 
applied to each McIlwain L interval considered in this analysis.

\subsection{Assessment of the systematic uncertainties}
\label{sec:sysStudy}
The main sources of systematic uncertainty in this analysis are:
the choice of the energy interval in which to fit the spectrum, the 
choice of the energy interval for the secondary template, 
the accuracy of the geomagnetic field model, and the accuracy of the
Monte Carlo simulations used for the estimation of the residual hadron 
contamination as well as the evaluation of the EGF.

To estimate the systematic uncertainty from the choice of energy interval 
used for the spectral fits, we performed a series of fits over a range 
of energy intervals narrower and broader than the chosen interval and 
calculated the root-mean square (rms) of the distribution of resulting values 
of the cutoff energy.  The rms was found to be no greater than $0.8\%$ 
for all the McIlwain L intervals.

Since the key ingredients in the calculation of the fraction 
of CRE primaries are the secondary and primary templates, it is crucial
to validate them. The tracer output (used as the template for the primaries)
can be validated directly with the Fermi LAT flight data, as was described 
in section~\ref{sec:tracer}. The secondary template is more difficult
to validate given that the output from the tracer code does not provide any
information regarding the secondary populations. As stated in section 
\ref{sec:primefrac}, it is safe to assume that the population at low energy 
in flight data (E$<<$E$_{\rm c}$) is predominately composed of secondary 
particles. However the choice of the energy interval is not well determined,
and can
therefore be a source of uncertainty. This choice can in fact influence
the shape of the spectrum and thus the final value for the cutoff rigidity.
To investigate its effect, we chose several different energies ranges in
which to define our secondary templates (for each McIlwain L interval) 
by varying both the width of the energy interval as well as the distance 
from the cutoff energy. With the resulting fraction of primaries obtained
from each template we reconstructed the spectrum and obtained the cutoff 
rigidity following the 
procedure described in section \ref{sec:cutoffrigidity}. The spread of the 
ratio of the cutoffs (found to be no greater than $\sim 2\%$ for all McIlwain L
intervals) gives an estimate for the uncertainty due to the secondary template.

The accuracy of the cutoffs derived from the particle tracing code is
limited by the uncertainty of the geomagnetic field model. Several
cross checks on the accuracy of the predicted rigidity cutoffs have been 
performed using satellite experiments, in particular
the cosmic-ray isotope experiment HEAO-3 C2~\cite{HEAO3C2} (C2) and
the Mass Spectrometer Telescope (MAST) on the Solar Anomalous and
Magnetic Particle Explorer (SAMPEX) spacecraft~\cite{SAMPEX}.
The HEAO-3 satellite flew between 1979 and 1981 at an 
altitude of 496 km and inclination of  
43.6$^{\circ}$. From the comparison between the experimental and computed 
cutoffs of oxygen nuclei in the 5 GeV/n range it was found that the calculated 
cutoffs were systematically $\sim$3--5 $\pm$ 2\% higher than the measured ones~\cite{HEAO3C2}. The SAMPEX mission was launched into an  
orbit with 82$^{\circ}$ inclination with an average altitude of $\sim$600 km and
operated from 1992 to mid 2004. They performed a comparison 
between measured cutoff rigidities of 0.3 to 1.7 GeV 
protons during geomagnetically quiet times with those predicted by the Smart
and Shea tracer code (based on IGRF-10) and found that the latter 
systematically overestimate the data by 8\%--14\%~\cite{SAMPEX2}. 

Directly comparing these findings with the measurements presented here is 
difficult because in both the SAMPEX and HEAO-3 cases the measured cutoff
rigidities are much lower than those appropriate for Fermi. Smart and Shea 
in~\cite{smartshea2} assert that the 
accuracy of the model improves in regions closer to the geomagnetic equator 
(i.e. for larger cutoff rigidities). According to their analysis, it is 
reasonable to assume that the bias in the tracer output used in this analysis 
is no greater than $\sim$3--5\% for the McIlwain L regions 
spanned by the Fermi observatory. However, this is just an extrapolation based 
on the measurements performed by HEAO-3 and SAMPEX at lower rigidities and
a definitive answer on the value of this bias at the Fermi rigidities 
remains unknown.

The uncertainty of the hadronic component of the Monte Carlo simulations used to
evaluate the residual contamination is estimated to be no greater than 20\%.
Its contribution to the overall systematic uncertainty in the final measurement
of the cutoff is small (0.5\%) because the contamination is
always below 20\% in every McIlwain L interval. The assessment of the systematic
uncertainty due to the EGF is described in detail in~\cite{fullpaper} and for 
this energy range has a value of 10--15\%. We studied its 
effect on the final value of the cutoff via a bracketing method. To do this we
generated a set of EGFs that have a maximal impact on the final 
shape of the spectrum, reconstructed the spectrum using these new EGFs and
analyzed the final value of the cutoff energy obtained with these new EGFs. The
resulting values varied by $\pm$1\%.

\subsection{Cross check of method}
To demonstrate that this analysis has the sensitivity to measure 
deviations of several percent in the energy scale of the LAT, we repeated 
the analysis on a test sample of traced particles with an added $+$5\%
shift in energy. The resulting measurement of the cutoff energy was 
5.4$\pm$ 0.2\% higher, which is consistent with the bias we introduced in 
the input. Thus we conclude that indeed this method is capable of measuring 
energy scale errors of the magnitude we expect.

Pre-launch tests of crystal boule samples showed~\cite{SaraCsIPaper} that the 
light yield of the CsI(Tl) crystals in the LAT calorimeter could be expected 
to decrease by $\sim$1\% per year from radiation damage in the charged-particle
environment of low Earth orbit, primarily from trapped particles in the South 
Atlantic Anomaly.

\begin{figure}[th!]
\begin{center}
\includegraphics[angle=90,width=\linewidth]{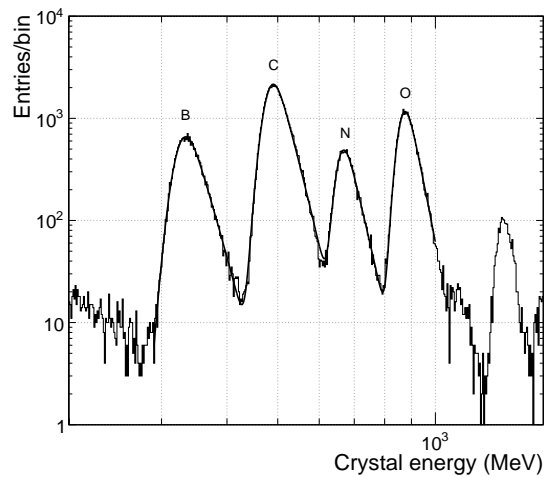}
\end{center}
\caption{The path length-corrected crystal energies of the B,C,N, and O 
peaks corresponding to the first four months of Fermi LAT operation. The 
large apparent abundance of B (compared to secondary and primary 
Galactic cosmic-ray abundance) is due to secondary B produced by primary C 
interacting in the ACD and TKR.}
\label{fig:BCNOfit}
\end{figure}

By analyzing the path length-corrected\footnote{The energy deposit is normalized by $\cos \theta$, where $\theta$ is the incidence angle of the particle.} 
crystal energies of four abundant GCR elements, namely boron, carbon, nitrogen 
and oxygen, over time
it is possible to verify this prediction with flight data. In fact, all 
four of these peaks 
show a similar linear decrease per year.
\begin{table}
\begin{center}
\begin{tabular}{cc}
\hline
& Drift (\%/yr)\\
\hline
B & 1.49$\pm$0.05\\
C & 1.60$\pm$0.04\\
N & 1.50$\pm$0.06\\
O & 1.46$\pm$0.03\\
\hline
CRE & 1.90$\pm$0.90\\
\hline
\end{tabular}
\end{center}
\caption{Drift per year for four abundant GRC element (B,C,N, and O) peaks 
and the CRE cutoff energy measured in Fermi LAT flight data over the first 
year of operations.}
\label{tab:tableElements}
\end{table}
These results are consistent with the pre-launch predictions. An example of 
these peaks is shown in figure \ref{fig:BCNOfit}. 
As a further cross check of the sensitivity of the method used in this analysis,
we have compared the cutoff rigidity values measured in the first 60 days 
to the last 60 days of the first year of operations. We found that the measured
cutoff energy decreased by 1.9$\pm$0.9\% over this time interval, which is 
consistent with the decrease from radiation damage measured with the GCR 
element peaks, as can be seen in table~\ref{tab:tableElements}.  We note also that this provides further confirmation of the 
sensitivity of our method.

\section{Results}

\begin{figure}[h!]
\begin{center}
\includegraphics[angle=90,width=\linewidth]{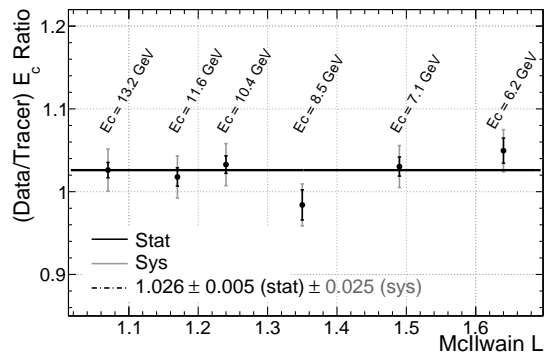}
\end{center}
\caption{Distribution of cutoff energy ratios as a function of McIlwain L. The
ratios are constant within the errors over the entire range. The quoted value 
for the ratio is the weighted mean. The statistical errors are represented 
by the black error bars, and the systematic errors are in gray. As a reference,
the value for the cutoff energy in each McIlwain L interval is also shown.}
\label{fig:calscale}
\end{figure}

We applied this analysis to a data set spanning the first year of LAT 
operations. We found that the measured cutoff energy exceeded the predicted 
cutoff energy by 1.026$\pm$0.005 (stat) $\pm$0.025 (sys) in the 6~GeV to 13~GeV
range. The systematic error is the sum in quadrature of our estimates in 
Section~\ref{sec:sysStudy}, excluding those from the uncertainty in the 
magnetic field model and particle tracing code.  As shown in 
Figure~\ref{fig:calscale}, the ratio between measured cutoff and predicted 
cutoff is constant within the errors over this energy range.

\section{Conclusions}

We provide a method for calibrating the LAT energy scale
based on the measurement of the geomagnetic cutoff in the primary CRE spectrum.
The results shown here confirm that the LAT calorimeter calibration
procedure, based on measured electronic gains and non-linearities, 
observed proton signals, and enforced consistency between readout energy 
ranges, is accurate to the $\sim$2\% level. It is worth noting that the
method developed here is the only such calibration which relies uniquely 
on flight data and uses 
electromagnetic showers in the LAT.  Accordingly, this is a very important 
cross check on other methods such as beam tests and calibrations with 
cosmic-ray heavy ions.

Our conclusion relies, however, on the fidelity of the current IGRF and 
particle trajectory tracing code. 
If we take the measurements of HEAO-3 C2 and
SAMPEX to indicate a bias of the order of +3-5\%, the absolute energy scale of 
the LAT may have to be decreased by 5--7\%.  This conclusion, if confirmed, 
would make the energy scale more consistent with the beam test results and the
Monte Carlo simulations of the CU. This would suggest that the on-orbit energy 
calibration technique for the LAT should be revised.
However, the cross checks reported in~\cite{HEAO3C2,SAMPEX2} on the
predictions of the tracer code were both performed where the cutoff 
rigidities are much lower than those measured by the LAT and
in geomagnetic regions where the trajectory tracing calculations are less 
accurate~\cite{smartshea}. It is therefore unclear
whether this bias in the theoretical cutoffs 
also applies to the results presented in this work. We therefore
prefer to determine the energy scale of the LAT using the results 
of the proton inter-range calibration, which are confirmed by the independent
measurements discussed in this paper.

\section{Acknowledgments}
The \textit{Fermi} LAT Collaboration acknowledges generous ongoing support
from a number of agencies and institutes that have supported both the
development and the operation of the LAT as well as scientific data analysis.
These include the National Aeronautics and Space Administration and the
Department of Energy in the United States, the Commissariat \`a l'Energie 
Atomique and the Centre National de la Recherche Scientifique / Institut 
National de Physique Nucl\'eaire et de Physique des Particules in France, the 
Agenzia Spaziale Italiana and the Istituto Nazionale di Fisica Nucleare in 
Italy, the Ministry of Education,
Culture, Sports, Science and Technology (MEXT), High Energy Accelerator Research
Organization (KEK) and Japan Aerospace Exploration Agency (JAXA) in Japan, and
the K.~A.~Wallenberg Foundation, the Swedish Research Council and the
Swedish National Space Board in Sweden.

Additional support for science analysis during the operations phase is 
gratefully acknowledged from the Istituto Nazionale di Astrofisica in Italy 
and the Centre National d'\'Etudes Spatiales in France.

\bibliographystyle{model1-num-names}
\biboptions{sort}
\bibliography{biblio_CalScale}

\end{document}

%% file: AuthorList.tex
\author[1]{M.~Ackermann}
\author[1]{M.~Ajello}
\author[1]{A.~Allafort}
\author[2]{W.~B.~Atwood}
\author[3,4,5]{M.~Axelsson}
\author[6]{L.~Baldini}
\author[7,8]{G.~Barbiellini}
\author[9,10]{D.~Bastieri}
\author[1]{K.~Bechtol}
\author[6]{R.~Bellazzini}
% ...
\address[1]{ W. W. Hansen Experimental Physics Laboratory, Kavli Institute for Particle Astrophysics and Cosmology, Department of Physics and SLAC National Accelerator Laboratory, Stanford University, Stanford, CA 94305, USA
}
\address[2]{ Santa Cruz Institute for Particle Physics, Department of Physics and Department of Astronomy and Astrophysics, University of California at Santa Cruz, Santa Cruz, CA 95064, USA
}
\address[3]{ Department of Astronomy, Stockholm University, SE-106 91 Stockholm, Sweden
}
\address[4]{ The Oskar Klein Centre for Cosmoparticle Physics, AlbaNova, SE-106 91 Stockholm, Sweden
}
\address[5]{ Department of Physics, Royal Institute of Technology (KTH), AlbaNova, SE-106 91 Stockholm, Sweden
}
\address[6]{ Istituto Nazionale di Fisica Nucleare, Sezione di Pisa, I-56127 Pisa, Italy
}
\address[7]{ Istituto Nazionale di Fisica Nucleare, Sezione di Trieste, I-34127 Trieste, Italy
}
\address[8]{ Dipartimento di Fisica, Universit\`a di Trieste, I-34127 Trieste, Italy
}
\address[9]{ Istituto Nazionale di Fisica Nucleare, Sezione di Padova, I-35131 Padova, Italy
}
\address[10]{ Dipartimento di Fisica ``G. Galilei", Universit\`a di Padova, I-35131 Padova, Italy
}
% ...
